\documentclass[aip,graphicx]{revtex4-1}
\usepackage{graphicx}
\usepackage[tight,TABTOPCAP]{subfigure}
\usepackage[]{SIunits}
\usepackage[usenames, dvipsnames]{color}
\usepackage{natbib}
\usepackage[normalem]{ulem}

\usepackage[utf8]{inputenc}
\usepackage{comment}

\draft 
\usepackage[utf8]{inputenc}
\usepackage[T1]{fontenc}
\usepackage{mathptmx}
\usepackage{etoolbox}

\makeatletter
\def\@email#1#2{%
 \endgroup
 \patchcmd{\titleblock@produce}
  {\frontmatter@RRAPformat}
  {\frontmatter@RRAPformat{\produce@RRAP{*#1\href{mailto:#2}{#2}}}\frontmatter@RRAPformat}
  {}{}
}%
\makeatother

\begin{document} 
\baselineskip24pt


\title{Dynamic Heterogeneity of Short Semi-crystalline Polymer Chains during Recrystallization} 

\author{Maziar Heidari}
\email{maziar.heidari@biophys.mpg.de}
\affiliation{Gulliver, CNRS UMR 7083, ESPCI Paris and PSL University, 75005 Paris, France}
\affiliation{Department of Theoretical Biophysics, Max Planck Institute of Biophysics, Max-von-Laue Straße 3, 60438, Frankfurt am Main, Germany}

\author{Matthieu Labousse}
\email{matthieu.labousse@espci.fr}
\affiliation{Gulliver, CNRS UMR 7083, ESPCI Paris and PSL University, 75005 Paris, France}

\author{Ludwik Leibler}
\affiliation{Gulliver, CNRS UMR 7083, ESPCI Paris and PSL University, 75005 Paris, France}
%
%
%


\begin{abstract}
The instant crystallization of semi-crystalline polymers have become possible following the recent advances in Fast Scanning Calorimetry (FSC) and enables to make a bridge between the time scale available experimentally with those accessible with computer simulations. Although the FSC observations have provided new information on the crystallization kinetics and evolution of the crystals, the molecular details on the chain exchange events between ordered and disordered domains of crystals have remained elusive. Using molecular dynamics simulations, we examined the detailed chain dynamics and thermodynamics of polyamide 6 (PA6) system under two heating treatments. (i) Quenching PA6 melt deeply below the melting temperature $T_m$ and (ii) annealing the resulting quenched system to a temperature close to $T_m$. We categorized the chains into mobile amorphous fraction (MAF) and rigid amorphous fraction (RAF), based on the length of consecutive chain's bond angles in trans state. In the deep quenched system close to the glass transition temperature $T_g$, the mobility of the MAF chains are strongly suppressed and they remain in glassy state. However, upon rising the temperature close to melting temperature, the system undergoes recrystallization leading to coexistence of RAF and supercooled liquid MAF chains. The highly mobile unentangled MAF chains explore the interphase domains, and during the late-stage of crystallization, they are thermally translocated into the lamellae by reducing the fold number of RAF chains. The chain mobility in the annealed system could potentially lead to improved biodegradation in semi-crystalline chains. 
\end{abstract}
\maketitle

\section*{Introduction}
The recent progress of experimental technology namely, chip sensor based fast scanning calorimetry, has opened the way to study fast crystallising systems ($>10^6\, \mathrm{K.s}^{-1}$) \cite{androsch2015density,furushima2017melting,toda2016insights} with time scales which become sufficiently small to be comparable to those accessible with computer simulations. Upon rapid quenching from melting temperature ($T_m$) to deep below the crystallization temperature ($T_c$), the crystallization starts with the homogeneous nucleation from the bulk melt. The nucleation process is followed by the further ordering of the folded chains to ultimate large-scale lamellae formation. In the interphase domain between neighboring lamellae, the chains that are partially contributing to the crystalline domain are categorized as rigid amorphous fractions (RAF) while for the chains completely inside the interphase disordered domain are named as mobile amorphous fractions (MAF). Inside the interphase region, the motion of the MAF chains are constrained by the entanglement of the other MAF chains and also by the disordered segments of the RAF chains. The mobility of the MAF chains is suppressed and glassy in the deep quenched close to the glass transition temperature ($T_g$). However, MAF chains become mobile and resemble supercooled melt for the temperatures close and below the melting temperature. Although recent studies \cite{androsch2015density,furushima2017melting,toda2016insights} focused on the change of MAF population in the recrystallization process as reflected in the dynamic heat capacity, the molecular details of transitions between RAF and MAF populations during recrystallization are not yet elusive. 

Alongside the long-time studies of polymer crystallization using experimental techniques \cite{strobl1980direct,reiter2007progress,reiter2001liquidlike,toda2016insights}, computer simulations have provided molecular details on the polymer crystallization which are difficult to probe in the experiments \cite{liu1998langevin,welch2001molecular,meyer2001formation,meyer2002formation,vettorel2007structural,jabbari2015plastic,ramos2015molecular,ramos2016coarse,flachmuller2021coarse}. Computer simulation of recrystallization of polymers was pioneered by Lu and Sommer \cite{luo2009coexistence} where they reported the 
coexistence of melting and growth of microcrystalline domains during heating of long (500-monomer) poly(vinyl alcohol) crystal. The associated large entanglement length of the long chains not only constraint the mobility of the MAF but it also lowers the relaxation time of locally ordered domains during the temperature changes, leading to memory effects \cite{xu2009cloning,luo2014frozen}.

In order to reduce the entanglement effect and investigate the behavior of the MAF chains during the late-stage of recrystallization, we performed computer simulations of short polyamide 6 (PA6) chains. There have been several computational models of PA at atomistic \cite{lukasheva2017influence} and coarse-grained resolutions \cite{carbone2008transferability,karimi2008fast,karimi2008hydrogen,eslami2013thick,eslami2011coarse}. Here, we used Martini representation of PA6 \cite{milani2011coarse} whose coarse-grained strategy is based on the polar and apolar categorizations of the PA monomer sub-units similar to protein residues \cite{marrink2007martini,monticelli2008martini}.
In the following, we investigate the MAF and RAF population dynamics and their exchanging events during quenching deeply below the crystallization temperature as well as reheating the quenched system close to the melting temperature. In the early stage of the recrystallization, the MAF chains adsorbed on the lamellae surface growing the size of crystal. However, in the later stage, the interphase region are explored by the unentangled MAF chains and they are translocated into lamellae via pushing the mult-folded RAF chains. The free energy change during the translocation is of order $k_BT$ which also assures the thermally driven mechanism.

\section*{Model}
We used a coarse-grained (CG) model of PA6 based on the Martini force field framework \cite{marrink2007martini,monticelli2008martini,milani2011coarse}. The CG model of PA6 consists of successive apolar and polar spherical beads which respectively, representing the alkane $-(\text{CH}_2)_4-$ and amide groups $-\text{CO}-\text{NH}-\text{CH}_2-$ (See Fig. S1a). We used LAMMPS simulation package to perform the simulations \cite{plimpton1995fast,thompson2022lammps}. Noser-Hoover thermostat was employed to control the temperature and pressure with damping coefficient of $10$ fs and $100$fs, respectively \cite{martyna1994constant,parrinello1981polymorphic,dullweber1997symplectic,shinoda2004rapid}. The pressure of the system was kept at 1 atm for all simulations. The equations of motion were integrated using Verlet algorithm with a time step of $1$ fs \cite{verlet1967computer,tuckerman2006liouville}. The non-bonded interactions were ruled by the Lennard-Jones (LJ) potential whose energy and length scales are given in the table S1. To verify the implemented force field, we simulated a short PA6 consisting of six monomers (12 units) with two n-butyl end groups and compared the distribution of bonds length and bond angles with the results provided in Milani {\it et al.}~\cite{milani2011coarse}. We used 916 chains of 24 monomers (48 CG units). The periodic boundary conditions were set along the three directions of space. Initially, the system was equilibrated at temperature $T=500$K and pressure $P=1$~atm for $22$~ns with the box side fluctuating about $15.72 \pm 0.01$~nm. Then the temperature of the systems was instantaneously changed into the target temperatures. 

 \begin{figure}[ht!]
    \centering
    \includegraphics[width=0.95\textwidth,angle=0]{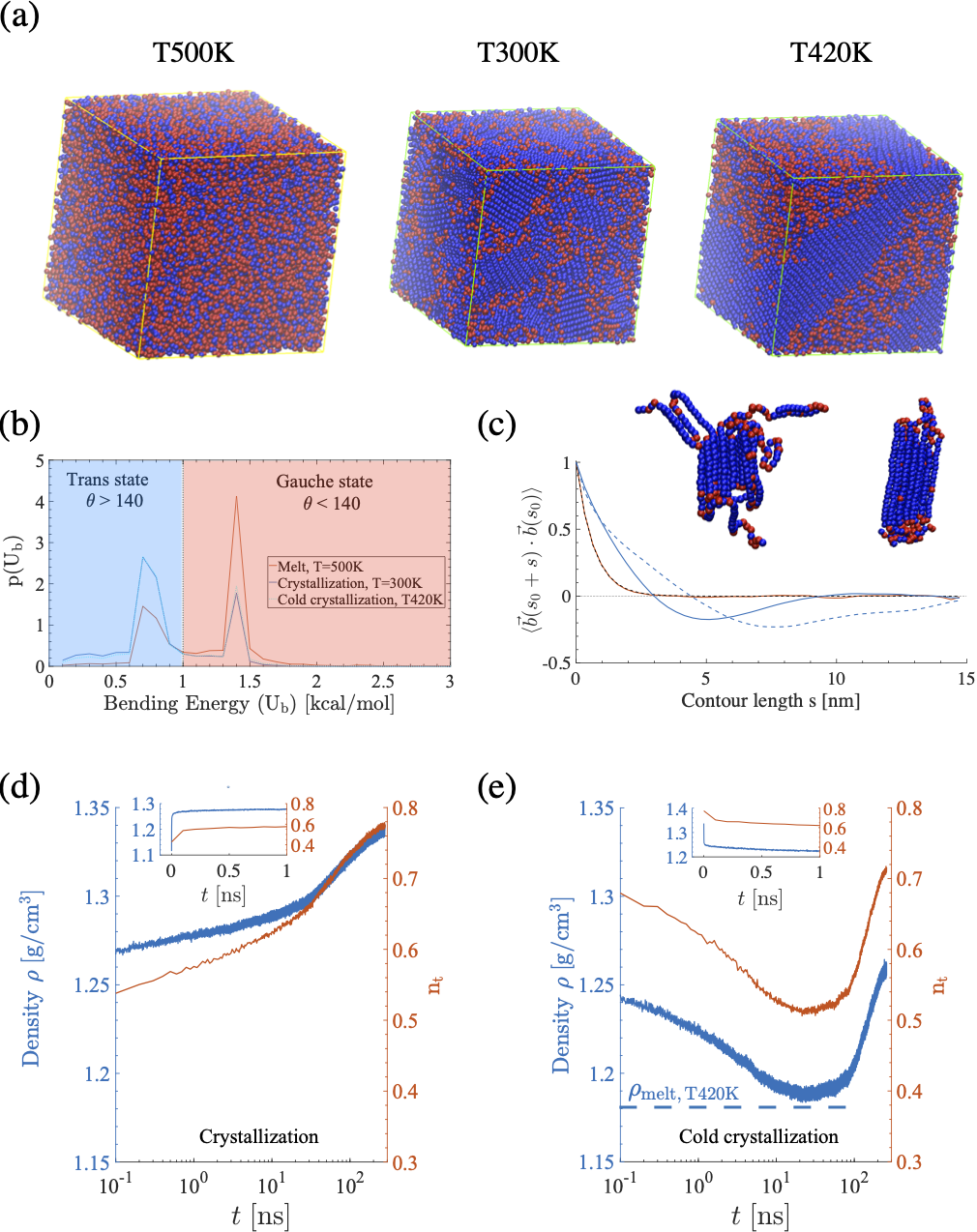}\\
\end{figure}
\addtocounter{figure}{0}

\begin{figure}[ht!]  
    \caption{Coarse-grained configuration of the polyamide 6 (PA6) chains at different temperatures. (a) Final snapshots of the polyamide chain configuration in melt state at T500K (\textit{left}), under quenching from T500K to T300K structure (\textit{middle}) and under recrystallization (cold crystallizing) from T300K to T420K (\textit{right}). The monomers are colored in red or blue if their corresponding bond angle is in gauche or trans state respectively (See panel (b) for the definition). (b) Probability density function of the bending energy of the PA6 angles at the end of simulations with temperature T500K (red curve), T300K (blue curve) and T420K (dashed blue). The angle is defined in trans state if the bending energy is below 1 kcal/mol or equivalently, the angle is above 140 degree and it is in gauche state otherwise. The system's density (blue curve) and fraction of the bond angles in trans state (red curve) are shown for quenched crystallization from T500K to T300K (d) and for re-crystallization from T300K to T420K (e). Insets in panel d and e show the initial ($0\le t\le 1$ns) values of density (blue curve) and trans population fraction (red curve). The horizontal dashed blue line in the panel e is the density of the supercooled PA6 melt at T420K.}\label{fig:Fig1}
\end{figure}
        
\section*{Results and discussion}

For the equilibrated molten state at T500K (this is the shorthand notation for $T=500K$, consistently used throughout the paper), the distribution of the normalized end-to-end distance $r=R_{ee}/\langle R_{ee}^2\rangle^{1/2}$ of PA6 chains follows the ideal chain distribution with slight deviation, $G_e(r)=4\pi r^2(3/2\pi)^{3/2}\exp{\left[-3r^2/2\right]}$, where the average end-to-end distance of the PA6 chains is $\langle R_{ee}\rangle=4.16$~nm (Fig. S2). This is in line with the Flory's argument on the ideal-chain conformational statistics of the chains in polymer melt \cite{flory1969statistical,hsu2016static}. Wittmer et al. \cite{wittmer2007polymer} has linked the deviation from the ideal chain distribution to the chain connectivity and melt incompressibility. In the following, we mainly focus on the two cooling and heating protocols, crystallization by quenching the PA6 melt from T500K to T300K and cold crystallization by heating the system quenched at T300K to T420K. In both protocols the temperature of the system is instantaneously changed. The Nose-Hoover thermostat and barostat respond to the changes according to their damping coefficients and maintain the temperature and pressure at the target values. Initially, equilibrated system at T500K is cooled down to T300K via different cooling rates. The evolution of specific volume ($\nu=1/\rho$) and nematic order parameter of the system (details of the nematic order parameter are explained in SI) under different cooling rates shows that the crystallization temperature is within T350K to T370K (Fig. S3). However, the crystallization of the polymers can occur in a wide temperature range \cite{xu2009cloning}. Also the PA6 melt can be in the meta-stable state (supercooled melt) for the temperature range larger than T370K, for example T420K, which can not be captured by our simulations \cite{heeley2003metastable} due to large free energy barrier between the supercooled melt state and crystalline state. Upon quenching to T300K (deep below the crystallization temperature), the system undergoes crystallization and lamellae formation. For each bond angle of the chains in the system, we defined the two states: trans state if the bending energy is lower than  1 $\text{kcal}.\text{mol}^{-1}$ and gauche state otherwise (Fig. \ref{fig:Fig1}b). The bending energy criterion corresponds to the angle threshold ($\theta =140^{\circ}$) which sets the energy barrier between two minima in the angle energy landscape (Fig. S1b). The configurations of the quenched (T300K) and cold-crystallized (T420K) systems at time 298 ns and 397 ns are shown in Fig. \ref{fig:Fig1}a. The monomers are colored in blue if the bond angle is in trans state and in red if they are in gauche state. In the melt (T500K), the chains are preferably populated in the gauche states (Fig. \ref{fig:Fig1}b) with fraction of $1-n_t=0.58$ (Inset of Fig. \ref{fig:Fig1}d). This is approximately identical to the Boltzmann weight obtained using $n_t=\mathcal{Z}^{-1}\int_{0^\circ}^{140^{\circ}}d\theta \exp\left[- U_b(\theta)\right/k_BT]\approx 0.4$ where $U_b(\theta)$ is the bending energy (Fig. S1b) and $\mathcal{Z}=\int_{0^\circ}^{180^\circ}d\theta \exp\left[- U_b(\theta)\right/k_BT]$ is the normalization factor.

The evolution of the density ($\rho$) and fraction of trans angles ($n_t$) are shown in Fig. \ref{fig:Fig1}d,e. 
During the initial crystallization ($t<50$~ns Fig. 1d), $n_t$ and $\rho$ rise simultaneously and there is no indication of preordering transition of the chains' segment prior to the crystallization \cite{imai1995ordering, matsuba1999conformational,olmsted1998spinodal}. The structure factors ($S(q)$) at different times from the onset of quenching is shown in Fig. S4a. During the initial quenching time ($t\le 10~\text{ns}$), the first Bragg peak $(1,0)$ which corresponds to the inverse of the monomer size ($\sigma^{-1}$) starts growing and shifts slightly toward higher wave vector. This is the signature of primary local packing between the neighboring monomers as the consequence of densification. As the crystallization continues, the Bragg peaks are sharpened and shifting toward higher wave vector. At the same time, the intensity of the $S(q)$ is enhanced for the wave vectors ranging between the lamellae size $l_{\rm lam}^{-1}\approx 1/5 ~\text{nm}^{-1}$ to the chain's end-to-end distance at T500K, 
 $R_{ee}^{-1}=1/4.11~\text{nm}^{-1}$. In the latest time of crystallization, second (1,1) and third (2,0) Bragg peaks emerge indicating 2D hexagonal lattice ordering \cite{vettorel2007structural}. The density of PA6 quenched at T300K reaches 1.35 [$\text{g}/\text{cm}^3$] which compares with $\alpha$ crystal phase of PA6 at T300K \cite{pauly1989polymer}. During reheating the system from T300K to T420K, two processes occurred sequentially. First the crystalline domains below the critical size start to melt and the density of the system drops to 1.183~$\text{g}/\text{cm}^3$ which is still above the meta-stable melt density at T420K (Dashed blue line in Fig. \ref{fig:Fig1}e shows the $\rho_{\text{melt}}^{\text{T420K}}=1.179~\text{g}/\text{cm}^3$ with standard deviation $0.001~ \text{g}/\text{cm}^3$). Second, the surviving crystalline domain starts to grow leading to rise in system density. The two sequential processes are also reflected in the structure factor analysis (Fig. S4b). The evolution of $n_t$ follows a similar trend as the density (Fig. \ref{fig:Fig1}e). We computed different energetic contribution of the system per monomer during crystallization and cold crystallization (Fig. S5). In crystallization (Fig. S5a), the bond energy rises as there are heterogeneous crystalline and amorphous regions in the system and the chains have to stretch to be adapted to the both regions. The bending energy and van der Waals energy decrease as there more angles in the crystalline region and the chains are more compact in the lamellae \cite{hall2019divining}. The two-step process in cold crystallization is also evident from the increase and successive decrease in the evolution of the system's energy (Fig. S5b).

The crystalline lamellae size is estimated by the bond vector correlation along the chain $\langle \vec{b}(s_0+s)\cdot \vec{b}(s_0)\rangle$. Here, the bond vector $\vec{b}$ is defined by the vector joining two successive monomer units along the chain and $s$ is the contour length. When the chains are in the melt state (T500K), the correlation decays exponentially $\sim e^{-s/l_p}$ giving the persistence length of $l_p=0.72$~nm. In the final snapshot of crystallined state and cold-crystallization state, the bond-bond correlation decreases more slowly than for the melt case and becomes negative for further segments (solid and dashed blue curve in Fig. \ref{fig:Fig1}d). Defining the minimum of the curve as the size of the lamellae, we obtained $l_{\rm lam}=5$~nm and $l_{\rm lam}=8$~nm for the quenched and cold crystallized lamellae. The corresponding lamellae of selected 9 chains are shown in the inset of Fig. \ref{fig:Fig1}c.

\begin{figure*}
    \centering
    \includegraphics[width=1.0\textwidth,angle=0]{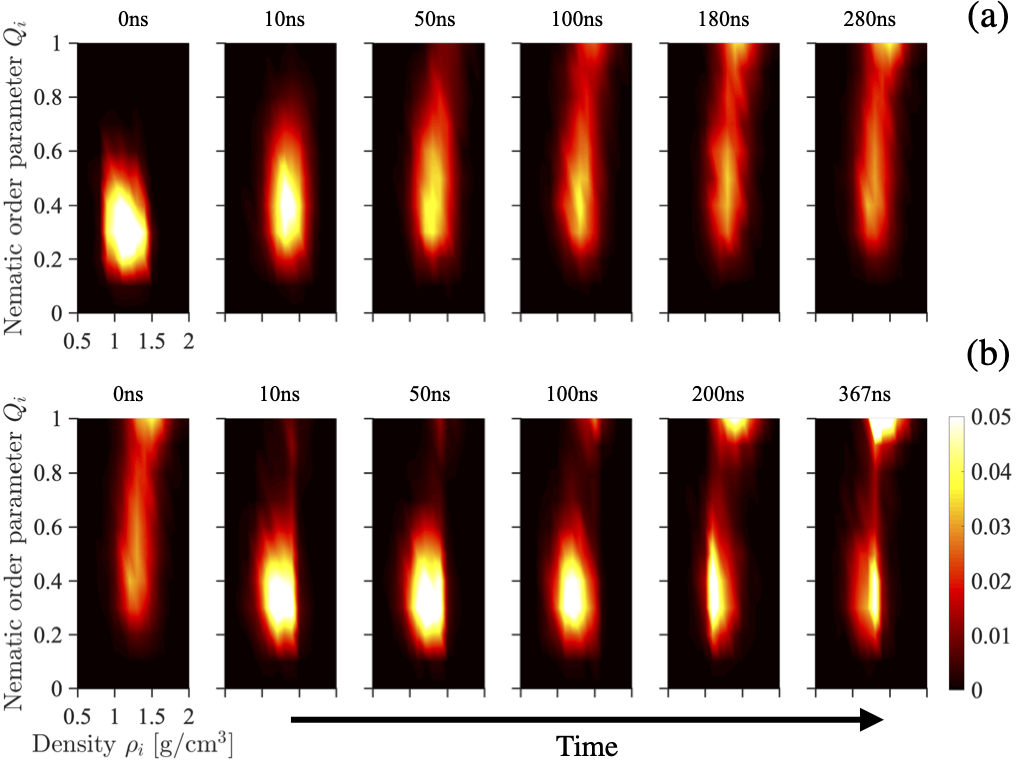}\\
    \caption{Joint probability distribution of the nematic order parameter and density of the system under crystallization and cold crystallization. The system is descritized into cubic box and in each box, the local density and nematic order parameter are calculated. (a) During quenching crystallization, at the t=0, the state of the melt at T500K is populated in a single peak and in the course of time, the system becomes heterogeneous in crystalline and amorphous regions. The crystalline regions contribute to the appearance of a new peak at higher density and higher nematic order parameter. (b) During cold crystallization, there exists two sequential processes, first the crystalline lamellae below the critical size melt while a single lamellae with large density and large nematic order parameter survive and starts growing. The population of the amorphous region also shrinks. }
    \label{fig:Fig2}
\end{figure*}

In order to investigate the local ordering and density of the two systems, we discretized the system into cubic voxels of size 1 nm, and for each voxel, we computed the local nematic order parameter $Q_i$ and the corresponding density $\rho_i$ (See SI for more details). The evolution of the $\rho_i-Q_i$ along crystallizarion and cold crystallization is shown in Fig.~\ref{fig:Fig2}. In the melt ($t=0$, Fig. \ref{fig:Fig2}a), the distribution of the local density shows a single peak with mean $\langle \rho_m\rangle=1.09~\text{g}.\text{cm}^{-3}$ and standard deviation $\sqrt{\langle \delta \rho^2 \rangle} = \sqrt{\langle (\rho - \langle \rho_m\rangle)^2\rangle} \approx 0.15$ $\text{g}.\text{cm}^{-3}$. The mean density is close to the experimental and numerical density of PA6 at $T=543$K reported in literature (0.96$\text{g}/\text{cm}^3$)~\cite{pauly1989polymer,milani2011coarse}. The isothermal compressibility of the PA6 melt at T500K is calculated using the local fluctuations of density inside the element volume $V$, $\chi_T= V\langle \delta \rho ^2\rangle/\rho^2 k_BT =2.68\times 10^{-9}~\text{Pa}^{-1}$ \cite{frenkel2023understanding, kardar2007statistical}. Upon quenching from T500K to T300K, the volume of the system suddenly shrinks. This consequences to translational shift of the mean density to higher density (see Fig. S6) as well as widening of $Q_i$ distribution. At the time $t=50~{\rm ns}$, as a result of lamellae formation, another peak appears at the high density and large local ordering parameter while the peak with lower density and order parameter corresponding to amorphous chains is still preserved. During the crystallization, ($t>50~{\rm ns}$), the population of the segments in the amorphous phase transits into crystalline phase. Toward the last simulation time ($t=280$~ns), the populations in the $\rho-Q$ map are barely changing leading to a broadened distribution between crystalline and amorphous domains. However, in the cold crystallization (Fig. \ref{fig:Fig2}b), upon increasing the temperature from T300K to T420K, the lamellae melts and the segments transit from crystalline region into amorphous melt $(10~{\rm ns} < t < 50~{\rm ns})$. However, due to the presence of stable lamellae, the peak corresponding to the crystalline domain does not completely vanish. In the course of simulation time, the lamellae starts grows as more chains from the amorphous region deposit on the lamellae. The growth process continues leading to more ordered crystalline phase compared with the quenched T300K case. The lamellae growth during the cold-crystallization are governed by two mechanisms: (\textit{i}) during the early stage, the lamella grows by adsorbing chains at the interface of the lamellae and amorphous region; (\textit{ii}) in the later stage, there are events during which tails of the chain protrude inside the lamellae forcing the tails of other folded chains to exit the lamellae. The nematic order parameter of the system which is sliced at three different planes (Fig. S7a) also shows the initial heterogeneously distributed lamellae at T300K and in the course of time, while a single locally ordered lamellae survives, the other lamellae melt. The resulting large lamellae at T420K melts via heating to T500K (Fig. S7b).

\begin{figure*}
    \centering
    \includegraphics[width=1.0\textwidth,angle=0]{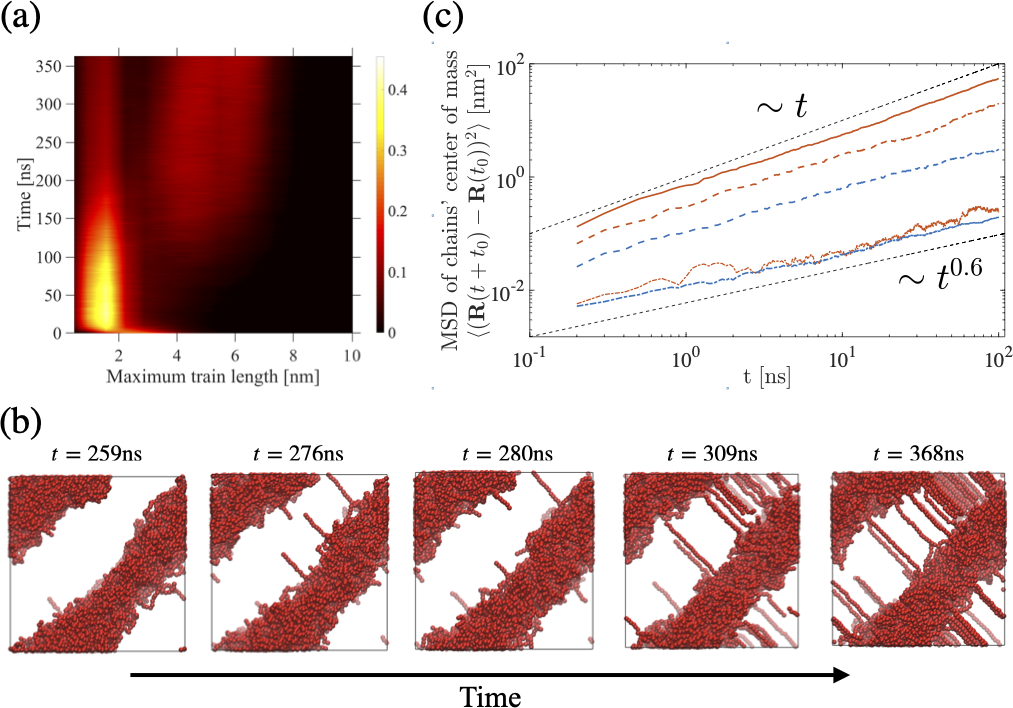}\\
    \caption{
    Avrami exponents and the time evolution of the extended volume fraction of crystalline domain during crystallization. The extended volume fraction of crystalline domain upon quenching from T500K to T300K (a) and reheating from T300K to T420K (b). The red line is the fitting curve to the simulation data. Here, we relate the extended volume fraction to the fraction of angles in trans state. For the quenched T=300K, we used the entire range ($t=0,\ldots ,300$~ns) and obtained $n=0.25$, $a=0.24$ and $b=0.55$. For T=420K, we excluded the initial melting regime ($t<10$ ns). There are clearly two intermediate and later regimes. We obtained $n=2.5$, $b=0.71$ and $a=10^{-6}$ for the range $10 {\rm ns}<t<150 {\rm ns}$ (dashed red line) and $n=0.41$, $b=0.052$ and $a=0.12$ for $t>150$ ns (dashed black line).}
    \label{fig:Fig3}
\end{figure*}

We used the so-called Avrami model (Kolmogorov-Johnson-Mel-Avrami) to investigate the time evolution of the crystallinity \cite{avrami1939kinetics,avrami1940kinetics}. Under the isothermal condition in which the nucleation and growth rates are constant, the dynamics of the crystallinity volume fraction $\phi(t)$ is expressed in terms of the extended volume fraction between the crystalline domains $\phi_0(t)$ through, $\phi_0(t)=\log(1-\phi(t)) =at^n$. Here, $a$ is a constant and $n$ is Avrami exponent. For our systems, we assume $\phi(t)$ to be equal to the fraction of bond angles in trans state, i.e. $\phi(t)=n_t(t)$. To account for the finite value of $n_t(0)\neq 0$ in the melt state, we used $\phi_0(t)=at^n+b$ to fit the data with $a$, $n$ and $b$ being the free parameters. For the quenched T300K, we obtained the exponent $n=0.25$ (Fig. \ref{fig:Fig3}a). While for re-crystallization process, we observed two different behaviors. For the initial stage of crystallization, $n=2.5$ while for the later stage $n=0.41$ (Fig. \ref{fig:Fig3}b). 
The results highlight the distinct crystallization and growth kinetics at low and high crystallization temperatures (Tc) \cite{toda2023crystal}. During isothermal recrystallization at a low temperature (300 K), the nuclei density is extremely high, leaving no room for crystal growth. In contrast, at a higher Tc (420 K), the nuclei density decreases, providing sufficient space for each nucleus to grow, making the process growth-dominated.

\begin{figure*}
    \centering
    \includegraphics[width=1.0\textwidth,angle=0]{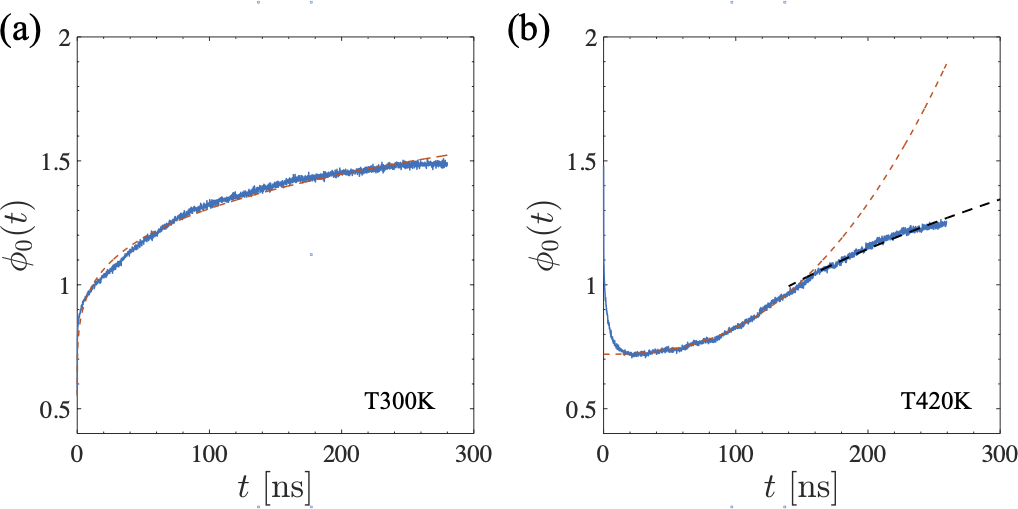}\\
    \caption{Dynamics of the maximum train length of the polyamide chains in the system. (a) The initial condition of the system is the quenched PA6 at T300K with the averaged maximum train length of $3.4$~nm (10 CG units). Then along the melting process, the chains leaves the crystalline phase and populates the crystalline phase. (b) Snapshot series of the MAF chains which are labeled at time t=259 ns of the simulation trajectory, is shown. The simulation time of each snapshot is labeled at the top of each panel. (c) The mean square displacement (MSD) of the chains' center of mass for PA6 melt at T500K (solid red line), MAF (dashed red line) and RAF and crystalline chains (dashed blue line) at T420K; and MAF (dash dotted red line) and RAF and crystalline chains (dash dotted blue line) at T300K. The dotted black lines show the diffusive ($\sim t$) and subdiffusive ($\sim t^{0.6}$) scaling. We set the reference position of the MSD of T420K and T300K at time $t_0=259~\mathrm{ns}$.}
    \label{fig:Fig4}
\end{figure*}

In order to investigate the collective chain dynamics during the crystallization, we plotted the distribution of the maximum train length of each chain as a function of time (Fig. \ref{fig:Fig4}a). For each folded chain, the maximum train length is defined as the maximum continuous length of the trans angles. At $t=0$ (T300K), the train length of the semi-crystalline system is distributed with the mean length $3.4$ nm (10 CG units). During the melting process (T420K), more chains contribute to the amorphous region with the mean value of the maximum train length being 2.2 nm. This length is larger than the persistence length of the chain at T500K ($l_p=0.72$ nm). Upon recrystallization, the population of the amorphous region mostly transits into the crystalline phase with an average length approximately $4.1$ nm to form a larger lamellae. The population exchange is also observed for other four independent simulation runs (Fig. S8-S11). In this regard, we defined two chain populations, mobile amorphous fraction (MAF) and crystalline chains which include the rigid amorphous fraction (RAF) and crystalline fractions. A chain is defined to be in crystalline population if its maximum train length exceed 5 CG units otherwise it is in amorphous phase. Fig.~\ref{fig:Fig4}b shows the sequence of snapshots of only MAF chains during the last 100 ns starting at t=258 ns (See also Movie 1). For the initial snapshot, we selected the MAF chains and colored them in red while the crystalline chains are not shown. At this stage in which the interfacial growth of lamellae has been almost completed, the mechanism of exchange of between the two population is via the replacement. The center of mass (CM) dynamics of the chains in each population is shown in Fig. \ref{fig:Fig4}c for the MAF and RAF populations at three different temperatures. There is clear dynamic heterogeneity in the MAF and RAF populations at T420K. Similar to PA6 melt at T500K (solid red line), the long-time CM dynamics of the amorphous chains at T420K are diffusive but with three times smaller diffusion constant ($D_{\rm T420K}\approx 0.03~\text{nm}^2/\text{ns}$ compared with $D_{\rm T500K}\approx 0.09~\text{nm}^2/\text{ns}$). The diffusion constants of the amorphous chains is also smaller than the diffusion constant of meta-stable PA6 melt at T420K due to the constraints and confinement imposed by RAF chains (See Fig. S12c). However, the dynamics of the crystalline chains (dashed blue line) which is mainly governed by the longitudinal motion inside the lamellae is much slower. The dynamic heterogeneity during the cold crystallization is more enhanced for a systems having a single lamella growth rather than multiple lamellae growth (see Fig. S11 and Fig. S12d). The translocation mechanism of the MAF chains is similar to permeation of molecules between layers of the smectic A liquid crystals \cite{mukherjee2013dual}. 

\begin{figure*}
    \centering
    \includegraphics[width=0.9\textwidth,angle=0]{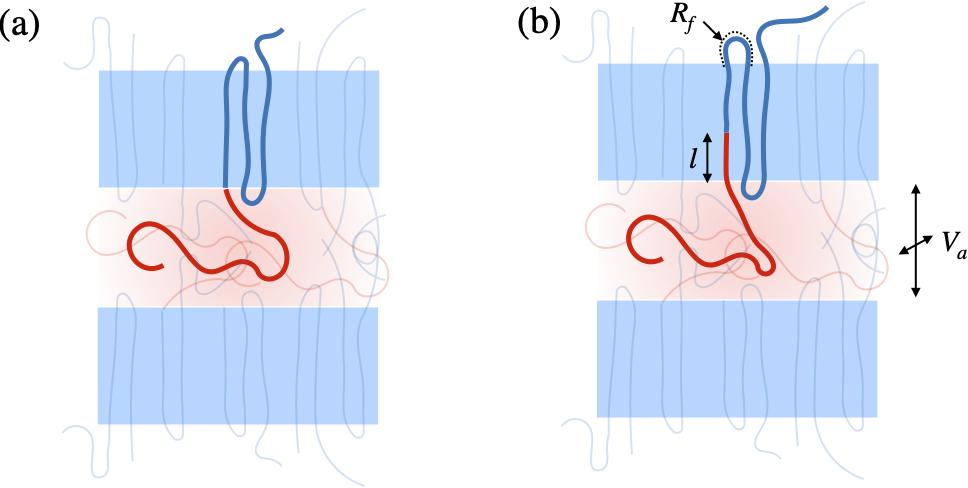}\\
    \caption{Chain replacement mechanism and unfolding of folded chains during the re-crystallization. Schematic of chain exchange during the re-crystallization of semi-crystalline polymers. The ordered crystalline and RAF chain population are shown in blue and MAF in red. In the late stage of re-crystallization, the amorphous chains are mobile within the interphase space (a) and they can translocate through the lamella by pushing multi-folded chains from the crystalline region (b). The interphase volume is denoted by $V_a$, the translocated length of the amorphous chain is $l$ and the radius of the curvature of the crystalline chain at the fold is $R_f$.}
    \label{fig:Fig5}
\end{figure*}

The schematics of the chain exchange between the amorphous and crystalline domains are shown in Fig. \ref{fig:Fig5} (See also Movie 2). The statistical mechanics theory on the amorphous chain fraction formed by loops and tails during crystallization have been thoroughly investigated by Sommer \cite{sommer2006role} and Muthukumar \cite{muthukumar2003molecular}. Here, we only look at the late-stage translocation of the MAF through the crystallite within the reservoir of MAF chains in the interphase region. Using the lattice chain model for the chain's configuration entropy \cite{de1979scaling}, the free energy change of the amorphous chain (red chain) upon translocation length of $l$ and monomer size $a$ is 
\begin{equation}
\Delta F_a=-\frac{l}{a}\Delta \epsilon +k_BT \left[\frac{l}{a} \log(z)+ \log\left(\frac{V_a}{n_f a^5}\right)\right]
\end{equation}

 where $\Delta \epsilon=\epsilon_c-\epsilon_a$ is the difference in the effective cohesive energy change of a monomer with other monomers between amorphous ($\epsilon_a$) and crystalline ($\epsilon_c$) domains. Here we neglect the roughness of the local potential surface inside the lamella. Upon translocating in the lamella, the chain loses its configurational and translational entropy. We used the entropy of an ideal chain on a lattice having $z$-coordination number and $V_a$ is available interphase volume and $n_f$ is the number density of at least single-folded RAF chains per unit area of the lamella surface. The interphase volume can be related to the total system's volume ($V_t$) and the system's crystallinity ($f_c$), via $V_a=(1-f_c)V_t$. During the translocation, we assume that the replacing tail of the crystalline chain reptates within the lamella which ultimately leading to single-fold loss. Depending on the distance of the re-entrance, the curvature energy of the release fold may change. Using the persistence length of the chain $l_p$ and the curvature of the fold $1/R_f$, the bending energy release via unfolding a single fold becomes $  (\pi l_p/2R_f) k_BT$ \cite{marantan2018mechanics}. Thus, the free energy change of the crystalline chain (blue chain) becomes 
\begin{equation}
\Delta F_c=\frac{l}{a}\Delta \epsilon - k_BT \frac{\pi l_p}{2R_f} -k_BT \frac{l}{a} \log(z) 
\end{equation}

The total free energy change per every translocation event and release of a single fold becomes
\begin{equation}
\Delta F_t = \Delta F_a+\Delta F_c= k_BT\left[ -\frac{\pi l_p}{2R_f}+\log\left(\frac{V_a}{n_fa^5}\right)\right]
\end{equation}

For our current PA6 model, we have the persistence length $l_p=0.72$ nm, radius of adjacent re-entry radius $R_f=0.47$ nm, system size $L=15.7$ nm, crystallinity fraction $f_c = 0.7$, and 2D hexagonal packing fraction of the chains inside the lamellae $\eta_{\mathrm{2D,Hex}}\approx 0.9$, with area number density $n_f=\eta_{\mathrm{2D,Hex}} L^2/\pi a^4$. We used $a=0.47~$nm as the monomer size. The total free energy change per every translocation event becomes $\Delta F_t\approx k_BT$ which is a small positive value, though tuning the parameters could also end in small negative value. An important implication is that the translocation events are thermally driven, which aligns with experimental observations of dynamic fold reduction in crystalline paraffin chains \cite{ungar1985crystallization}.

\section*{Conclusion}
Using a coarse-grained model of PA6, we investigated the dynamics of the MAF and RAF chains during two heat treatment pathways: quenching PA6 melt from T500K to below crystallization temperature T300K and reheating crystalline PA6 from T300K to T420K. Based on the maximum train length of bond angles in trans state, we defined two chain populations, amorphous (MAF) and crystalline (RAF and crystalline fraction) chains. At the early stage of re-crystallization, the lamella start to melt and the chains mostly populate the amorphous state. At the intermediate stage, the amorphous chains contribute to the interfacial lamella growth. In the late stage of re-crystallization, the mechanism of exchange between the two population becomes different. MAF chains start to translocate into the lamella via their ends and replace the two or three folded crystalline chains' ends. We show that the free energy change upon exchange depends on the bending stiffness and the volume of the interphase between the lamellae. The center of mass dynamics of the two populations are clearly heterogeneous; MAF chains are mobile and exhibit long-time diffusive dynamics with diffusion coefficient less than the corresponding meta-stable melt while RAF chains posses subdiffusive dynamics. Our findings are important for understanding the semi-crystalline polymers during the heating treatment and in particular for the polymer hydrolase and degradation as the enzymes are mostly interacting with disordered region \cite{tournier2020engineered, wei2022mechanism,yamashita2005enzymatic,gaillard2019experimental,gomes2022lessons}.

\section*{Supporting Information Available}
\begin{itemize}
\itemsep-1em 
    \item[] Supporting Materials
    \item[] Supporting Table S1
    \item[] Supporting Movies S1 to S2
    \item[] Supporting Figures S1 to S12
\end{itemize}
\bibliography{bib_file}

\end{document}